\documentclass[conference]{IEEEtran}
\IEEEoverridecommandlockouts
\usepackage{marvosym}
\usepackage{url}

\usepackage{amsmath,amssymb,amsfonts}
\usepackage{algorithm}
\usepackage{algpseudocode} 

\usepackage{graphicx}
\usepackage{textcomp}
\usepackage{xcolor}
\usepackage{booktabs}
\usepackage{subcaption}
\usepackage{multirow}
\usepackage{tabularx}

\def\BibTeX{{\rm B\kern-.05em{\sc i\kern-.025em b}\kern-.08em
    T\kern-.1667em\lower.7ex\hbox{E}\kern-.125emX}}
\begin{document}

\title{GasLiteAA: Optimizing ERC-4337 for Efficient and Secure Gas Sponsorship}

\author{
\IEEEauthorblockN{
    \begin{tabular}{ccc}
        Hongxu Su\IEEEauthorrefmark{1}\textsuperscript{*}\thanks{\textsuperscript{*}Equal contribution. \textsuperscript{\Letter}Corresponding author.} & 
        Mingzhe Liu\IEEEauthorrefmark{2}\textsuperscript{*} & 
        Jie Xu\IEEEauthorrefmark{2} \\
        \textit{hsu238@connect.hkust-gz.edu.cn} & 
        \textit{mingzliu5-c@my.cityu.edu.hk} & 
        \textit{jiexu49-c@my.cityu.edu.hk} \\[0.1cm]
    \end{tabular}
    \\
    \begin{tabular}{cc}
        Xiaohua Jia\IEEEauthorrefmark{2} & 
        Xuechao Wang\IEEEauthorrefmark{1}\textsuperscript{\Letter}\\
        \textit{csjia@cityu.edu.hk} & 
        \textit{xuechaowang@hkust-gz.edu.cn} \\
    \end{tabular}
}

\vspace{0.1cm}

\IEEEauthorblockA{
    \IEEEauthorrefmark{1}\textit{The Hong Kong University of Science and Technology (Guangzhou)}, Guangzhou, China \\
    \IEEEauthorrefmark{2}\textit{City University of Hong Kong}, Hong Kong, China
    }
}

\maketitle

\begin{abstract}
ERC-4337, the Ethereum account abstraction standard, simplifies account management and transaction fee payment in decentralized applications by introducing programmable smart contract wallets and gas sponsorship via paymasters. However, its heavy reliance on on-chain validation and frequent state updates incurs substantial gas overhead, leading to performance bottlenecks and limiting scalability in large-scale deployments.
To mitigate these issues, we propose \textit{GasLiteAA}, a framework that optimize ERC-4337 by offloading paymaster logic to Trusted Execution Environments (TEE). GasLiteAA delegates the secure execution of stateful gas sponsorship logic and user quota management to TEE, enforcing validation rules off-chain while anchoring their integrity on-chain via lightweight cryptographic attestations. This verifiable offloading architecture significantly reduces on-chain computation and storage costs without sacrificing verifiability or decentralization.
Experimental results demonstrate that GasLiteAA substantially lowers transaction fees, while remaining fully compatible with Ethereum Layer~1. By balancing security, efficiency, and deployability, GasLiteAA provides a practical and scalable approach to gas sponsorship for account-abstraction–based decentralized applications.
\end{abstract}


\section{Introduction} \label{sec:introduction}

\noindent \textbf{Challenges in Ethereum account management.} In recent years, Ethereum~\cite{buterin2013ethereum} has established itself as a leading platform for decentralized applications, enabling innovations across various sectors. However, its broader adoption remains limited due to the complexity of wallet management~\cite{sheridan2022web3}. Externally Owned Accounts (EOAs) are the default user identities on Ethereum, whose control depends entirely on a corresponding private key. However, this authentication model creates a fragile user experience due to key loss, theft, and the inability to execute automated transactions~\cite{c1,9461136,praitheeshan2019attainable}. These challenges are especially severe in high-frequency transaction scenarios, reducing usability and hindering application growth~\cite{carter2021defi,werner2022sok}. Consequently, improving wallet usability and security has become a critical step towards enabling Ethereum's mainstream adoption~\cite{moniruzzaman2020examining, Kumar}. 

\noindent \textbf{Account Abstraction and ERC-4337.} Account Abstraction (AA)~\cite{AA} is a paradigm that generalizes the concept of accounts on Ethereum by enabling smart contract-based accounts~\cite{Elrom2019} to perform the same operations as EOAs, thereby decoupling transaction logic from authentication mechanisms~\cite{Wang_2023, Jiao_2024}. Introduced to address the usability and security challenges of EOAs, AA allows for advanced functionalities such as programmable spending limits, social recovery, and multi-factor authentication, significantly improving the user experience~\cite{wu2022time}. Several major wallet providers, including MetaMask~\cite{Metamask}, Trust Wallet~\cite{TrustWallet}, and Argent~\cite{Argent}, have launched AA wallets, while Ethereum-compatible platforms like Polygon, BNB Chain, and zkSync have also integrated AA. According to Dune Analytics~\cite{Dune}, the cumulative number of AA-related User Operations (\textit{userOps}) exceeded 193.25 million as of 2025.

ERC-4337~\cite{eip4337_1} offers a practical and decentralized framework for implementing AA without requiring modifications to the core Ethereum protocol. This allows ERC-4337 to be deployed on Ethereum mainnet and compatible blockchains without compromising stability or backward compatibility. However, its reliance on interconnected components such as \textit{Bundlers}, \textit{EntryPoint} contracts, and \textit{Paymasters} also increases system complexity~\cite{Lin_2024_02}. Among them, the \textit{Paymaster} increases the flexibility of gas fee management, making gas sponsorship more practical across different use cases~\cite{Andrews_2023_08}. By simplifying transaction fee management, \textit{Paymasters} lower the barrier to entry for users. Existing projects such as Biconomy~\cite{Biconomy}, OpenGSN~\cite{OpenGSN}, and StackUp~\cite{Stackup} have implemented \textit{Paymaster} functionalities, offering platforms that streamline user interactions and improve usability. 

\noindent \textbf{Computational bottlenecks and ERC-4337 optimization challenges.} While the \textit{Paymaster} mechanism provides flexibility in managing transaction fees, it introduces additional computational overhead due to the validation logic required for secure and abuse-resistant operations. For instance, a simple daily gas limit check adds approximately 5,360 gas per transaction, representing a 25.5\% increase over standard transfers~\cite{standardgas}. These costs escalate drastically with more complex logic such as dynamic limits or multi-factor authorization, increasing transaction fees and reducing network throughput. To address this inefficiency, offloading execution emerges as a promising pathway to mitigate gas costs, allowing computations to be performed off-chain with results submitted to the blockchain for verification. However, it also introduces a new set of critical challenges. Moving validation logic off-chain risks compromising the system's \textbf{compatibility}, \textbf{transparency}, and \textbf{trustworthiness}. This motivates the need for a solution that improves performance without sacrificing Ethereum’s decentralized security guarantees.

\noindent \textbf{Our solution.} To address these challenges, we propose GasLiteAA, an optimized framework for ERC-4337 that utilizes Trusted Execution Environments (TEE)~\cite{TEE} to offload complex \textit{Paymaster} validation logic. Unlike rollup-based approaches that introduce an additional execution layer, GasLiteAA specifically targets the computational bottleneck of gas sponsorship while retaining full integration with the Ethereum Layer 1 (L1) mainnet. By shifting state transitions to secure TEE enclaves, our framework reduces on-chain overhead without changing ERC-4337’s core architecture or compromising permissionlessness. Crucially, to ensure integrity and auditability, GasLiteAA generates verifiable cryptographic attestations within the TEE, certifying the correctness of off-chain computations. These attestations are anchored on-chain, enabling transparent validation of operations without relying on centralized intermediaries.

\noindent \textbf{Contributions.} Our main contributions are as follows:
\begin{itemize}
    \item We propose GasLiteAA, a framework that leverages TEE to optimize ERC-4337 \textit{Paymaster} validation, enabling efficient gas sponsorship without sacrificing verifiability or decentralization.
    
    \item We design a comprehensive system including deterministic routing for \textit{Bundler} coordination, Merkle tree-based state commitment, and an economic incentive mechanism with staking and slashing to ensure execution consistency and system integrity.
    
    \item We implement a robust prototype (28k LoC) and evaluate its performance comprehensively. Extensive experiments demonstrate up to 48.2\% gas reduction compared to on-chain verification, along with a 9,300$\times$ reduction in latency and a 2,200$\times$ reduction in memory usage compared to Zero Knowledge Proof (ZKP) alternatives.
\end{itemize}
\section{Background and Related Work}
\subsection{Background}
\label{sec:Background}
\textbf{Ethereum's growth highlights the need for AA.} 
Despite Ethereum’s role as the dominant settlement layer, the rigidity of traditional EOAs remains a major bottleneck to mainstream adoption on L1.
Even in 2025, the user experience is plagued by the ``single point of failure'' risk associated with private keys. Security reports indicate that phishing attacks and ``blind signing'' incidents on the mainnet continue to result in billions of dollars in asset losses annually. Personal wallet compromises now represent a growing share of total ecosystem theft, with attackers increasingly targeting individual users, making up 23.35\% of all stolen-funds activity YTD in 2025. This issue is inherent in the limited capabilities of EOAs~\cite{chainalysis_2025_crime}. 
The industry's decisive shift towards AA is evidenced by the recent inclusion of EIP-7702 in the Pectra upgrade~\cite{eip7702}. This protocol change allows EOAs on L1 to temporarily authorize smart contract logic, signaling a shift toward programmable accounts in Ethereum interaction. GasLiteAA can further synergize with EIP-7702 by improving the efficiency of programmable-account transactions, thereby accelerating the broader adoption of account programmability. 
Furthermore, as L1 evolves into a hub for high-value DeFi and Intent-based architectures, the need for atomic batch transactions, session keys, and social recovery has never been greater. AA bridges the gap between complex blockchain protocols and user safety, turning the mainnet into a more secure and user-centric environment compatible with modern authentication methods such as Passkeys~\cite{webauthn}.

\noindent \textbf{AA redefines transaction validation on Ethereum.}
AA redefines transaction validation on Ethereum by allowing developers to specify custom validity conditions, called ``programmable transaction validity''. Unlike Ethereum’s traditional model requiring fixed validation rules (e.g., gas balance and signature checks), AA enables flexible rule customization via Smart Contract Wallets (\textit{SCWs}) and \textit{userOps}~\cite{di2020wallet, di2020characteristics, tavares2018wallid}. A notable implementation of AA at the L1 level is ERC-4337, which provides a decentralized and backward-compatible framework for introducing these advanced functionalities without requiring changes to Ethereum’s core protocol. 

\noindent \textbf{ERC-4337 Components and Execution Flow.} To realize Account Abstraction, ERC-4337 introduces a framework centered around five key components: \textit{userOps}, \textit{Bundlers}, the \textit{EntryPoint}, \textit{SCWs}, and \textit{Paymasters}. The process begins with a user generating a \textit{userOp} (a data structure describing the transaction intent) and submitting it to a \textit{Bundler}. The \textit{Bundler} performs initial validation and broadcasts the \textit{userOp} to a decentralized \textit{mempool}. Subsequently, the \textit{Bundler} aggregates multiple operations into a batch and submits them to the \textit{EntryPoint} contract, a singleton hub that coordinates the execution. In this flow, the \textit{EntryPoint} ensures security by verifying signatures, nonces, and gas sponsorship eligibility through the \textit{SCW} and \textit{Paymaster}.

\noindent \textbf{Transaction Execution and Validation.} Once a batch is submitted, the \textit{EntryPoint} triggers the on-chain validation logic. It first verifies the \textit{userOp} through the \textit{SCW} and then consults the \textit{Paymaster} to determine if it will sponsor the gas fees based on predefined rules. If the sponsorship is approved and sufficient gas is pre-deposited, the \textit{userOp} is executed, and the \textit{Bundler} is compensated by either the user or the sponsor. In the current implementation of AA, the \textit{Paymaster} and its validation logic are entirely executed on-chain to ensure transparency. However, this approach introduces significant overhead, as the computational and storage costs associated with on-chain validation directly increase gas fees. These additional costs are ultimately borne by the user or the \textit{Paymaster}, highlighting a critical limitation of the existing framework.

\subsection{Related Work}
The existing \textit{Paymaster} verification mechanisms can primarily be categorized into two types: traditional on-chain verification and off-chain signature-based verification. Both mechanisms play a crucial role in the implementation of AA, offering different advantages and challenges.

\noindent \textbf{On-chain Paymaster verification.}
The standard approach, on-chain verification, ensures that all sponsorship logic—such as checking contract states or managing user quotas—is executed directly on the blockchain. This method is the foundation of the ERC-4337 reference implementation by \textit{Eth-Infinitism}~\cite{infinitism} and is adopted by leading infrastructure providers like \textit{Safe}~\cite{safe2025glossary} and \textit{Pimlico}~\cite{pimlico}. For instance, \textit{Pimlico}'s ERC-20 \textit{Paymaster} enables users to pay gas fees in tokens like USDC by performing the entire currency exchange and validation process on-chain to ensure transparency. Furthermore, Layer 2 networks with native AA support, such as \textit{zkSync}~\cite{zksync} and \textit{Starknet}~\cite{starknet}, as well as ecosystem-led initiatives like \textit{Base's} gasless campaigns~\cite{coinbasePaymaster}, rely on on-chain logic to maintain permissionless fee management. While this model guarantees auditability and security through decentralized consensus, it incurs significant computational overhead. Every validation rule and state update increases the gas cost per \textit{userOp}, creating a critical scalability bottleneck for high-frequency or complex sponsorship scenarios.

\noindent \textbf{Off-chain Paymaster verification by signature.}
An alternative approach is the off-chain signature model, often implemented as a ``Verifying Paymaster.'' In this model, users submit transaction details to an off-chain service provided by platforms such as Pimlico~\cite{pimlico}, Alchemy~\cite{alchemyGasManager}, or Biconomy~\cite{biconomyPaymaster}. The service evaluates sponsorship eligibility based on private or complex business logic and returns a cryptographic signature to be included in the \textit{userOp}. The on-chain \textit{Paymaster} then merely verifies this signature, which significantly reduces gas consumption compared with executing the full validation logic on-chain. While highly efficient, this mechanism introduces significant centralization and transparency concerns. The off-chain signer acts as a ``black box'' gatekeeper with unchecked authority to approve or deny requests. This lack of auditability can lead to censorship or unfairness, as users cannot verify the decision-making process, undermining Ethereum’s decentralized ethos. 

\section{Problem Statement and System Model} 
\subsection{Problem Statement}
While off-chain execution resolves the computational bottlenecks highlighted in Section~\ref{sec:introduction}, it disrupts the native trust model of ERC-4337. To optimize performance without sacrificing decentralization or security, the following critical challenges must be resolved: 

\subsubsection{\textbf{Compatibility with ERC-4337}} While aiming to optimize ERC-4337 by offloading part of the computational burden to off-chain execution, it is critical to ensure seamless compatibility with Ethereum's L1 networks. A viable solution must remain fully integrated with the Ethereum mainnet, enabling direct interaction with existing decentralized applications (DApps). This requires avoiding separate execution layers so that GasLiteAA can be adopted across the Ethereum ecosystem without disrupting existing workflows. 

\subsubsection{\textbf{Decentralization and Auditability}} Offloading the on-chain \textit{Paymaster} validation logic to off-chain execution is not a novel concept. Some existing solutions adopt a model where users submit sponsorship requests directly to sponsors, who then use their off-chain data to decide whether to approve the transaction. If approved, the sponsor provides a signed message that the user submits on-chain to facilitate payment by the \textit{Paymaster}. While this approach reduces on-chain overhead, it raises significant concerns about centralization and transparency. Since the entire off-chain verification process is fully controlled by the sponsor, users lack the ability to verify the fairness or correctness of the sponsor’s decisions. This centralization compromises auditability and ultimately undermines user trust and the integrity of the system.

\subsubsection{\textbf{Execution Consistency}} Unlike rollups that operate on independent state roots, ERC-4337 \textit{userOps} are executed directly on L1. A critical misalignment arises when off-chain validation approves a \textit{userOp} and updates the off-chain \textit{UserState}, but the subsequent on-chain execution fails. In such cases, the on-chain state remains unchanged while the off-chain state effectively drifts, leading to inconsistencies. A robust framework must therefore guarantee atomicity, ensuring that the commitment to the updated \textit{UserState} is finalized on-chain if and only if the corresponding L1 execution succeeds.
    
\subsubsection{\textbf{Bundler Coordination}} For censorship resistance, it is preferable to have multiple Bundlers working in parallel. However, this can lead to independent processing of \textit{userOps}, causing conflicting \textit{UserState} updates. Different \textit{Bundlers} may process different \textit{userOps}, causing off-chain state divergence and conflicting on-chain updates that threaten system integrity.

\subsection{System Model} \label{sec:system_model}

We define \emph{allocation rules} as policies governing how a \textit{Paymaster} allocates gas resources to users. Each rule is indexed by $r_j$ and specifies the logic for validating and updating gas sponsorship requests. 
Let $u_i$ denote a unique user, $t$ denote a discrete time index (e.g., a daily epoch or block height), and $c$ represent the gas cost of a pending operation.

Under rule $r_j$, the specific accounting data required to validate user $u_i$ at time $t$ is referred to as the \textit{UserState}, denoted by $S_t^{(r_j)}(u_i)$. 
Rather than imposing implementation-specific constraints, we treat $S_t^{(r_j)}(u_i)$ as an abstract state that captures all context needed to enforce the allocation policy (e.g., remaining quotas or historical usage).

Efficient verification is achieved by committing these \textit{UserStates} off-chain into a Merkle tree. The leaves of this tree represent the hashed state of individual users, while the resulting root, denoted by $T_{\mathrm{root}}^{(r_j,b_i)}$, serves as a cryptographic commitment to the aggregate state processed by bundler $b_i$ under rule $r_j$. This root is the only application-level state data stored on-chain, explicitly distinguishing application-level \textit{UserStates} from Ethereum’s native \textit{AccountState}.

GasLiteAA integrates the system model described above into a practical framework by enhancing the standard ERC-4337 workflow with TEE-based off-chain computation. By shifting the burden of \textit{Paymaster} rule verification to secure enclaves, the system maintains the auditability of \textit{UserState} transitions while minimizing on-chain costs. The architecture is composed of the following additional modules: 

\begin{figure*}[htbp]
    \centering
    \includegraphics[width=0.9\textwidth]{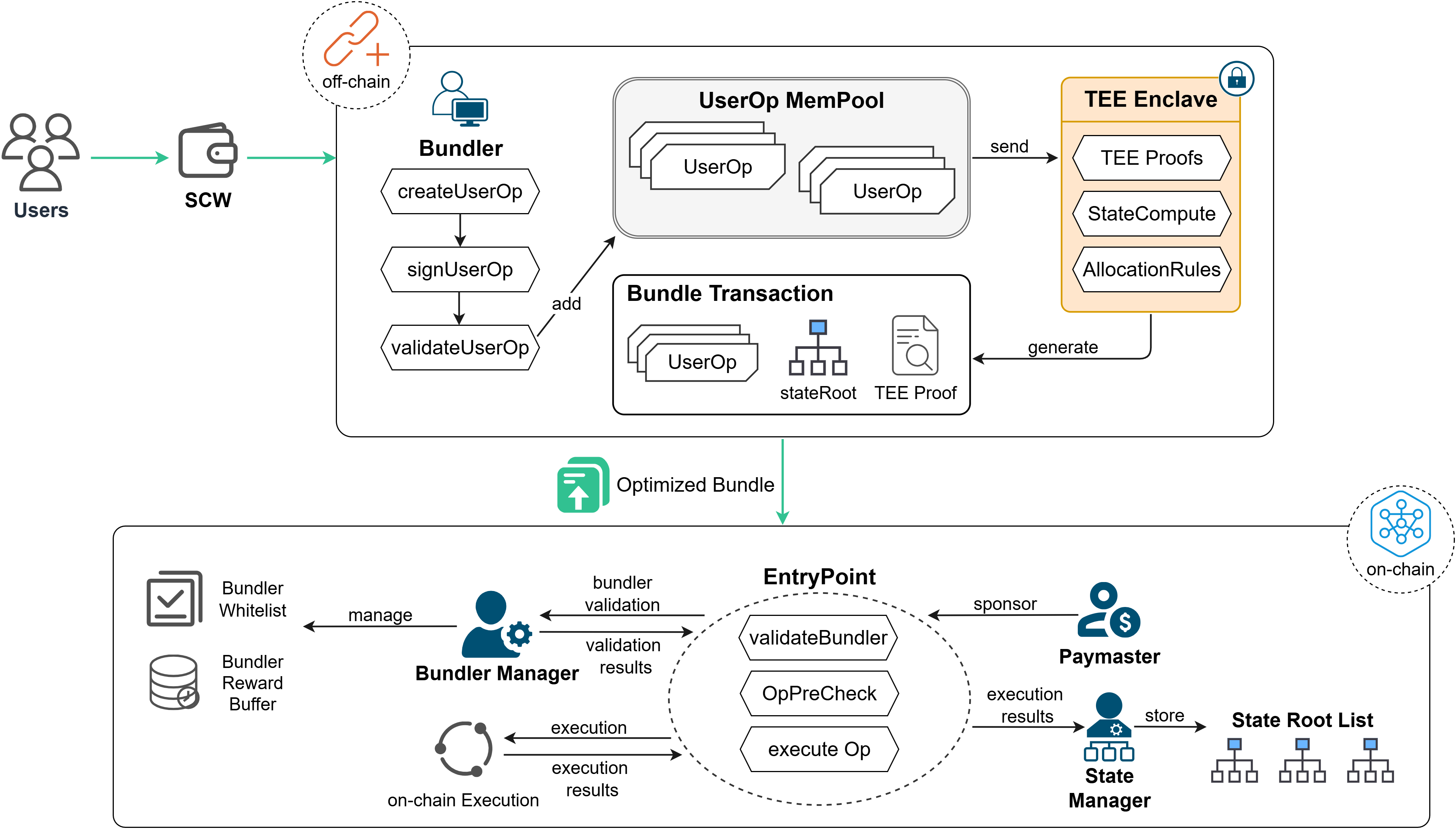}
    \caption{GasLiteAA Framework Overview.}
    \label{fig:GasLiteAA_overview}
\end{figure*}

\begin{itemize}
    \item \textbf{Bundler Manager \& Reward Buffer}: A system governing \textit{Bundler} participation through refundable stakes and a delayed reward pool. This setup deters malicious behavior via slashing and allows for recovery in case of state inconsistencies.
    \item \textbf{State Manager}: Responsible for maintaining the \textit{UserStateRoot} and its history on-chain to ensure verifiable commitments and error recovery.
    \item \textbf{TEE Enclave}: A secure off-chain environment where allocation rules are executed confidentially. It generates cryptographic proofs to certify the correctness of state transitions.
    \item \textbf{AttestationOnChainVerifier}: An on-chain component that validates TEE proofs. Successful verification ensures that off-chain calculations are strictly bound to on-chain state updates.
\end{itemize}

\section{Proposed Framework}

\subsection{GasLiteAA Overview}
As shown in Figure~\ref{fig:GasLiteAA_overview}, GasLiteAA uses a hybrid workflow that combines secure off-chain execution with verifiable on-chain settlement. The process begins with users submitting \textit{userOps} to the mempool, where a deterministic routing mechanism assigns them to designated \textit{Bundlers}. Utilizing a TEE, the \textit{Bundler} calculates state transitions based on the \textit{UserState} definitions (Section~\ref{sec:system_model}) and predefined allocation rules. This off-chain computation produces an updated \textit{UserStateRoot} and a corresponding TEE Proof. In the subsequent on-chain phase, the \textit{EntryPoint} executes the transactions and anchors the new state root to Layer 1. This architecture effectively offloads computational overhead while preserving decentralization and security. The following subsections detail the sequential steps of this workflow. 

\subsection{Detailed System Design}

\subsubsection{Sending and Routing \textit{userOps}}
To ensure fairness and prevent synchronization issues caused by multiple \textit{Bundlers} competing for the same \textit{UserState}, GasLiteAA employs a deterministic routing mechanism. Using the notation defined in Section~\ref{sec:system_model}, the system assigns each \textit{userOp} to a specific bundler index $b_i$ based on the user identity $u_i$, the current block index $t_{\mathrm{block}}$, and the active bundlers set $B$:
\begin{align}\label{eq:bundler_index}
b_i=\bigl(\text{keccak256}(u_i) + t_{\mathrm{block}}\bigr)
\bmod |B| .
\end{align}
This mechanism ensures intra-block consistency and cyclic redistribution across subsequent blocks, guaranteeing eventual processing and discouraging ``lazy'' \textit{Bundler} behavior. The \textit{Bundler Manager} maintains the active set $B$ via a circular indexing system, seamlessly redistributing responsibilities during \textit{Bundler} churn to preserve continuity. Users can broadcast operations to the general mempool or utilize wallet-assisted routing to reach the designated \textit{Bundler}. Finally, Block Builders verify this assignment during validation to maintain censorship resistance.

\subsubsection{Off-chain Allocation Execution}
Upon receiving the \textit{userOps}, the \textit{Bundler} manages off-chain execution through a sophisticated approach integrated with the on-chain state to ensure consistency.

\noindent \textbf{Bundle Preparation and Pre-validation.} 
The GasLiteAA \textit{Bundler} first aggregates a batch of pending  \textit{userOps} from the mempool based on the routing protocol. Before invoking the TEE, it performs standard off-chain simulations (e.g., signature and nonce checks) to filter invalid requests and prevent resource exhaustion within the enclave.  Once a candidate bundle $\mathcal{B}$ is formed, it is prepared as input for the secure enclave. 

\noindent \textbf{State Transition in TEE Enclaves.} 
Subsequently, the \textit{Bundler} extracts the allocation rule specified in each \textit{userOp} and retrieves the latest Merkle proof for the \textit{UserState} referencing the on-chain \textit{UserStateRoot}. Within the TEE enclave, the bundle $\mathcal{B}$ and the current user state $S_t$ are used to compute the updated state $S_{t+1}$ via a state transition function $f(S_t, \mathcal{B})$. 

The enclave commits these updates by recomputing the Merkle Tree Root. The updated leaf for a user $u_i$ under rule $r_j$ is defined as the hash of the user identity and their new allocation state:
\begin{align} \label{eq:state_leaf}
    \mathrm{Leaf}(u_i, r_j) = \operatorname{Hash}(u_i \parallel S_{t+1}^{(r_j)}(u_i)).
\end{align}
The enclave then recursively hashes the updated leaves to obtain the new root node $T_{\mathrm{root}}^{(r_j, b_i)}$. This root represents the aggregate state for all users managed by \textit{Bundler} $b_i$. 

To guarantee integrity, the TEE Enclave signs $T_{\text{root}}^{(r_j, b_i)}$ using its private key $k_{\text{TEE}}$, producing a signature
\begin{align}\label{eq:signature}
    \sigma = \operatorname{Sign}_{k_{\mathrm{TEE}}}
\bigl(T_{\mathrm{root}}^{(r_j, b_i)}\bigr).
\end{align}
This cryptographic proof ensures the computation is tamper-resistant. Once completed, the signed results are submitted to the \textit{EntryPoint} contract for on-chain verification and execution.

\noindent \textbf{Submit Optimized Bundle.} 
Finally, the \textit{Bundler} \( b_i \) submits the execution results to the \textit{EntryPoint}. 
To ensure atomic verification, these results are encapsulated into an optimized bundle structure, denoted as $\mathrm{Bundle}_{\mathrm{opt}}^{(b_i,r_j)}$. 
As defined in Eq.~(\ref{eq:bundle_optimized}), this structure aggregates the original batch of \textit{userOps} ($\mathrm{Bundle}$), the updated \textit{UserStateRoot} \( T_{\mathrm{root}}^{(r_j,b_i)} \) computed within the TEE, and the cryptographic signature $\sigma$ attesting to the validity of the state transition: 

\begin{align}\label{eq:bundle_optimized}
\mathrm{Bundle}_{\mathrm{opt}}^{(b_i,r_j)}
\;=\;
\Bigl(
\mathrm{Bundle},\;
T_{\mathrm{root}}^{(r_j,b_i)},\;
\sigma
\Bigr).
\end{align}

\subsubsection{On-chain Execution and State Anchoring}
In the on-chain process, the \textit{EntryPoint} first verifies the legitimacy of the submitting \textit{Bundler} via the \textit{Bundler Manager}. Upon successful verification, the \textit{EntryPoint} executes the \textit{userOps} and ensures that the \textit{Paymaster} covers the associated gas fees. The \textit{Bundler}’s rewards are then accrued in the \textit{Bundler Reward Buffer}, providing an auditable and unified system for reward management. Simultaneously, the \textit{UserStateRoot} is updated in the \textit{State Manager}, and the associated TEE Proof is anchored on-chain to guarantee transparency and verifiability. The update of the \textit{UserStateRoot} is atomic with the transaction execution. While \textit{Bundlers} utilize pre-execution simulations to minimize failures, any on-chain revert automatically invalidates the proposed state transition, preventing state drift. For recovery, the full \textit{UserState} and its ordered update log are persisted on InterPlanetary File System (IPFS), while only the committed root is maintained on-chain. If a \textit{Bundler} becomes unavailable or an on-chain execution reverts, the affected \textit{userOps} are returned to the mempool for reprocessing, and the valid state can be deterministically reconstructed from the last confirmed root and the ordered log.

\subsubsection{Verifying Bundler's Integrity}
In GasLiteAA, the integrity of \textit{Bundlers} is enforced through a combination of economic staking and on-chain cryptographic verification. 
\textit{Bundlers} are required to lock collateral in the \textit{Bundler Reward Buffer}, creating a direct economic deterrent against malicious behavior. 
At the technical level, the \textit{AttestationOnChainVerifier} validates the correctness of off-chain execution. 
This validation logic is formalized as $\operatorname{VerifyBundle}$, which mandates the simultaneous satisfaction of two conditions: verifying state consistency against the current on-chain root $T_{\text{root}}^{(r_j,b_i)}$, and authenticating the validity of the proposed new root $T_{\text{root,new}}^{(r_j,b_i)}$ via cryptographic signature, as defined in Eq.~(\ref{eq:verify_bundle}):

\begin{multline}\label{eq:verify_bundle}
\operatorname{VerifyBundle}
=
\Bigl\{
\operatorname{VerifyRoot}\!\left(
\mathrm{Bundle},\;
T_{\mathrm{root}}^{(r_j,b_i)}
\right)
\Bigr\}
\\
\wedge\;
\Bigl\{
\operatorname{VerifySig}\!\left(
\sigma,\;
T_{\mathrm{root,new}}^{(r_j,b_i)},\;
k_{\mathrm{TEE}}^{\mathrm{pub}}
\right)
\Bigr\}.
\end{multline}

If this verification fails, or if verifiable misbehavior is reported, the \textit{Bundler Manager} initiates a slashing procedure, forfeiting the \textit{Bundler}'s stake and rewarding the reporter. 
Together, these mechanisms ensure accountability, security, and transparency. Even if the enclave or its signing key $k_{\mathrm{TEE}}$ is compromised, the impact is confined to \textit{Paymaster}-level gas allocation and \textit{UserStateRoot} forgery. User assets remain protected because every \textit{userOp} must still pass independent on-chain ECDSA signature verification by the \textit{SCW}.

\section{Implementation and Evaluation}
This section evaluates GasLiteAA by benchmarking its performance, scalability, and cost-effectiveness against two representative paradigms: (1) \textbf{Native On-chain Verification} (the ERC-4337 standard), which represents the maximum decentralization but is constrained by EVM gas limits; and (2) \textbf{ZKP Verification}, a leading cryptographic alternative for off-chain verifiable computation. While ZKPs offer trustless verification similar to TEE, they differ significantly in computational intensity and proof generation time. By implementing all three approaches under identical allocation rules, we aim to quantify the trade-offs between TEEs, ZKPs, and traditional on-chain execution.

\subsection{Implementation}
In our implementation, the core components, including the \textit{EntryPoint}, SCWs, and \textit{Paymaster}, were deployed and tested in multiple environments. During local testing, we deployed the core smart contracts with \textbf{Hardhat} to simulate \textit{userOps} and smart wallet interactions, validating contract logic before testnet deployment. Our local environment was configured with Node.js v19.9.0 and Hardhat v2.19.4. After local testing, the core smart contracts were deployed on the Sepolia testnet to verify interactions under real blockchain conditions. 

The \textit {Bundler} and \textit {userOp} mempool components were deployed on a Microsoft Azure Hyper-V virtual machine with an Intel Xeon Platinum 8370C CPU (2.80GHz, up to 3.49GHz), 32 GiB of memory, and Intel SGX support. The \textit{Bundler}'s StateUpdate module runs within an Intel SGX enclave, providing tamper-proof execution. Additionally, during \textit{Bundler} execution, an SGX Attestation Report is generated and uploaded to a smart contract on the Sepolia testnet for integrity verification. By combining local testing, testnet deployment, and TEE-based cloud execution, our implementation ensures both functional correctness and security for \textit{userOps}.

For the implementation of routing and validation logic, the \textit{Bundler} requests task allocation by signing the current block number to authenticate with the server-hosted mempool. Upon receiving the batch, the \textit{Bundler} verifies the ownership of each SCW via standard ECDSA signatures. It then parses the \texttt{paymasterAndData} field from each \textit{userOp} to identify and inject the specific allocation parameters into the TEE enclave. Finally, the resulting execution proofs and updated state roots are packaged and submitted to the \textit{EntryPoint} smart contract by invoking the standard \texttt{handleOps} function. To formally summarize this hybrid workflow, Algorithm~\ref{alg:bundler_process} outlines the precise execution logic.

\begin{algorithm}[ht]
\caption{Bundler Execution Flow with TEE Integration}
\label{alg:bundler_process}
\begin{algorithmic}[1]
\Require Batch of userOps $\mathcal{B}_{in}$; Current on-chain root $T_{\text{root,old}}$; Active Rules $\{r_j\}$
\Ensure Submission payload to \textit{EntryPoint}

\State \textbf{Phase 1: Host Pre-validation}
\State $\mathcal{B}_{valid} \gets \emptyset$
\ForAll{$op \in \mathcal{B}_{in}$}
    \If{$\text{SimulateValidation}(op) == \text{success}$}
        \State $\mathcal{B}_{valid}.\text{add}(op)$
    \EndIf
\EndFor

\State \textbf{Phase 2: TEE Execution}
\State $(T_{\text{root,new}}, \; \sigma) \gets \Call{TEE\_Execute}{\mathcal{B}_{valid}, T_{\text{root,old}}}$

\State \textbf{Phase 3: On-chain Submission}
\State Call \texttt{handleOps}$(\mathcal{B}_{valid}, \; T_{\text{root,new}}, \sigma)$ on \textit{EntryPoint}

\Statex
\Procedure{TEE\_Execute}{$\mathcal{B}, \; T_{\mathrm{root}}$}
    \State Load state $S$ from $T_{\mathrm{root}}$
    \ForAll{$op \in \mathcal{B}$}
        \State $r_j \gets \text{ParseRule}(op.\text{paymasterAndData})$
        \State $u_i \gets op.\text{sender}$
        \State $S^{(r_j)}(u_i) \gets \text{ApplyRule}(S^{(r_j)}(u_i), op)$
    \EndFor
    \State $T_{\text{root,new}} \gets \text{ComputeMerkleRoot}(S)$
    \State $\sigma \gets \text{Sign}_{k_{\text{TEE}}}(T_{\text{root,new}})$
    \State \Return $(T_{\text{root,new}}, \sigma)$
\EndProcedure
\end{algorithmic}
\end{algorithm}

The \textit{EntryPoint} contract safeguards the integrity of off-chain computations by invoking the \textit{AttestationOnChainVerifier}, a component based on Puffer Finance’s RAVE framework~\cite{PufferFinanceRAVE}. It uses the \texttt{verify\_epid()} function to validate the submitted \textit{userReportData} and \textit{encodedAttestation}. It checks the enclave measurement (\texttt{MRENCLAVE}), the freshness of the attestation report, and the binding between the enclave identity and the attested public key.
This validation process incurs an additional gas cost of approximately 60,000. Once verified, the \textit{userOps} are processed and committed to the blockchain, ensuring that the updates adhere to the allocation rules and maintaining system consistency. 

To establish a comprehensive benchmark, we developed an alternative implementation of the allocation rules using ZoKrates and an on-chain ZK verifier. We selected ZoKrates because of its mature toolchain for generating EVM-compatible Solidity verifiers from high-level circuits. The allocation rules were compiled into ZK circuits to evaluate the trade-offs between privacy, security, and efficiency. Specifically, we used ZoKrates to compute the witness based on allocation constraints, generate a proof ensuring computation correctness, and export a verifier for on-chain validation. We conducted comparative tests to evaluate the feasibility of using ZKPs versus TEE. These tests measured the end-to-end latency and on-chain gas consumption, providing insights into the performance differences.

Our evaluations show that the ZK verification function \texttt{verifyTx()} requires about 4$\times$ as much gas as the \texttt{verify\_epid()} function used for TEE attestation.  Moreover, the total latency for the ZK approach, which includes witness generation, proof computation, and verification, is substantially greater than that of the TEE-based workflow. This indicates that while ZKPs provide trustless verification, they incur higher on-chain cost and end-to-end latency than GasLiteAA. These performance disparities and scalability trade-offs are examined in detail in the following evaluation.

\subsection{Evaluation}
We evaluated GasLiteAA using 1000 SCWs across three distinct scenarios: (1) \textbf{Infinitism (On-chain)}: the traditional model where rules are verified on the EVM; (2) \textbf{ZK-Baseline}: where rules generate a ZKP off-chain for on-chain verification; and (3) \textbf{GasLiteAA}: our TEE-optimized framework. 

\noindent \textbf{Experimental Setup.}
For the experiments, we measured the gas consumption, execution time, and resource overhead while processing varying batches of \textit{userOps}, specifically: 1, 20, 50, 100, 200, 400, 600, 800, and 1000. 

Consistent with the implementation setup, the off-chain components for GasLiteAA (TEE) were hosted on the \textit{Microsoft Azure Hyper-V virtual machine} equipped with an \textit{Intel Xeon Platinum 8370C CPU (SGX-enabled)} and 32 GiB of RAM. For a fair comparison, the ZKP generation was performed on the same hardware environment.

The allocation rules used in the tests range in complexity (see Tables~\ref{tab:gas_comparison_simple} and \ref{tab:gas_complexity_comparison}), defined by user $u_i$, current gas cost $c$, cumulative usage $U(u_i)$, and limit $L$:

\begin{itemize}
\item \textbf{Rule 1: Fixed Daily Gas Cost Allocation}: Limits a user's cumulative daily gas $U_{\text{day}}(u_i)$ plus current cost $c$ against a daily limit $L_{\text{daily}}$:
\[
U_{\text{day}}(u_i) + c \le L_{\text{daily}}
\]

\item \textbf{Rule 2: Fixed Daily and Global Cap}: Adds a global constraint where $G_{\text{day}}$ is the total gas consumed by all users:
\[
U_{\text{day}}(u_i) + c \le L_{\text{daily}}, \quad G_{\text{day}} + c \le L_{\text{total}}
\]

\item \textbf{Rule 3: Frequency and Interval}: Enforces a minimum interval $\Delta t$ since the last operation time $t_{\text{last}}$, a single-op limit $L_{\text{one}}$, and a window quota $U_{\text{win}}$:
\[
t_{\text{last}}(u_i) + \Delta t < t_{\text{block}}, \quad c \le L_{\text{one}}, \quad U_{\text{win}}(u_i) + c \le L_{\text{win}}
\]

\item \textbf{Rule 4: Dynamic Daily Limit}: Adjusts the limit $L_{\text{dyn}}$ based on the wallet balance $\beta(u_i)$ and historical average gas $\overline{H}(u_i)$:
\[
L_{\text{dyn}} = L_{\text{base}} + \frac{\beta(u_i)}{10\,\text{ETH}} - \frac{\overline{H}(u_i)}{10}
\]
\end{itemize}

\subsubsection{Gas Consumption and Verification Overhead}
We first analyzed the performance under high load conditions (1000 \textit{userOps}) using the most complex allocation logic (Rule 4). As shown in Table~\ref{tab:gas_comparison_simple}, the traditional on-chain approach (Infinitism) incurs a prohibitive gas cost of 158.48 million. 

\begin{table}[htbp]
\caption{Gas Usage: GasLiteAA vs. Infinitism (Rule 4)}
\label{tab:gas_comparison_simple}
\centering
\begin{tabular}{cccc}
\toprule
\textbf{\begin{tabular}[c]{@{}c@{}}Batch Size\\ (UserOps)\end{tabular}} & \textbf{\begin{tabular}[c]{@{}c@{}}GasLiteAA\\ (Total Gas)\end{tabular}} & \textbf{\begin{tabular}[c]{@{}c@{}}Infinitism\\ (Total Gas)\end{tabular}} & \textbf{\begin{tabular}[c]{@{}c@{}}Overhead\\ (Infinitism)\end{tabular}} \\ \midrule
1 & 501 k & 566 k & +13.0\% \\
20 & 2.04 M & 3.08 M & +51.0\% \\
50 & 4.48 M & 7.07 M & +57.7\% \\
100 & 8.55 M & 13.71 M & +60.4\% \\
200 & 16.72 M & 31.89 M & +90.7\% \\
400 & 33.23 M & 63.75 M & +91.9\% \\
600 & 49.34 M & 95.09 M & +92.7\% \\
800 & 65.72 M & 126.78 M & +92.9\% \\
\textbf{1000} & \textbf{82.15 M} & \textbf{158.48 M} & \textbf{+92.9\%} \\ \bottomrule
\end{tabular}%
\end{table}

In contrast, both off-chain approaches (GasLiteAA and ZK) significantly reduce the total gas cost to approximately 82 million. It is important to note that the base execution gas for the \textit{userOps} remains constant across both off-chain approaches. Consequently, the marginal difference in total gas cost (82.15 M vs. 82.33 M) stems entirely from the distinct on-chain verification mechanisms: GasLiteAA requires only a constant and lightweight signature check ($\approx$ 60,000 gas), whereas the ZK approach incurs a higher cost ($\approx$ 235,154 gas) for verifying the cryptographic proof on-chain. 

Table~\ref{tab:gas_comparison_simple} compares gas costs across batch sizes under complex logic. GasLiteAA demonstrates significantly better scalability, maintaining a linear gas growth. In contrast, the Infinitism approach suffers from substantial overhead, which nearly doubles the gas cost (+92.9\%) at high loads (1000 \textit{userOps}) due to expensive on-chain verification operations.
\begin{figure}[t]
    \centering
    \includegraphics[width=\linewidth]{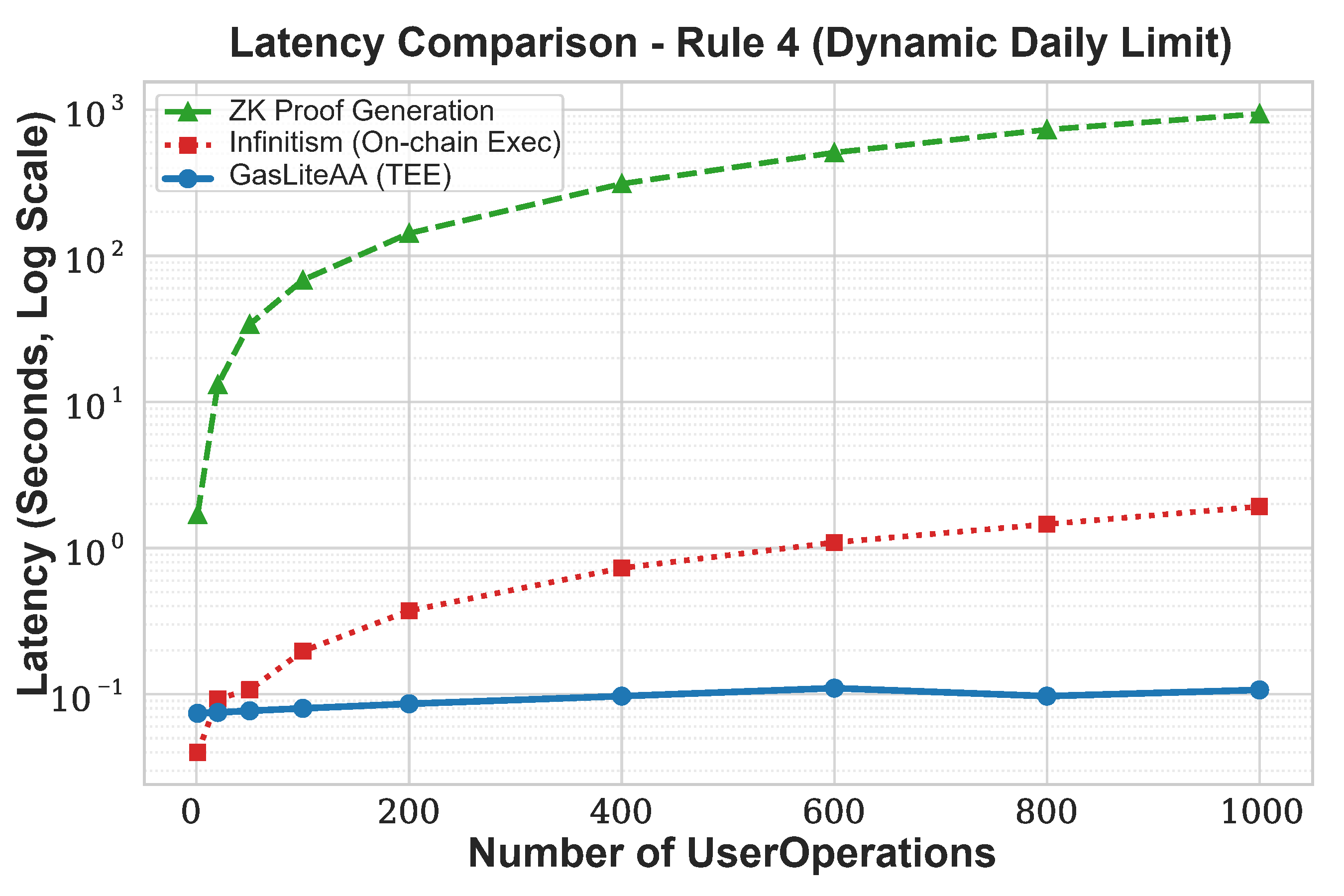}
    \caption{Off-chain computation latency comparison.}
    \label{fig:latency_all_rules}
\end{figure}

\subsubsection{Scalability and Resource Constraints}
While ZK and TEE approaches offer similar gas savings, their scalability limits differ fundamentally due to distinct computational bottlenecks. For the traditional Infinitism framework, scalability is severely constrained by the Ethereum block gas limit (30 million). As noted in previous comparisons, a batch of 1000 \textit{userOps} requiring 158.48 million gas is impossible to settle in a single block, creating a hard ceiling on throughput.

The ZK approach solves this on-chain gas limitation but introduces a new off-chain scalability hurdle: resource intensity. As detailed in Table~\ref{tab:zk_bottleneck}, the computational overhead for ZK proofs grows significantly with complexity. For complex rules (Rule 4), the memory requirement surges to over 26 GB. This excessive resource demand hinders horizontal prover scaling without prohibitive hardware costs, whereas GasLiteAA maintains a minimal footprint.

\begin{table}[htbp]
\caption{ZK Resource Bottleneck Analysis at 1000 \textit{userOps} across Rules}
\label{tab:zk_bottleneck}
\centering
\resizebox{\columnwidth}{!}{%
\begin{tabular}{@{}lccccc@{}}
\toprule
\textbf{Rule Type} & \textbf{\begin{tabular}[c]{@{}c@{}}Time\\ (s)\end{tabular}} & \textbf{\begin{tabular}[c]{@{}c@{}}Memory\\ (MB)\end{tabular}} & \textbf{\begin{tabular}[c]{@{}c@{}}Circuit Size\\ (MB)\end{tabular}} & \textbf{\begin{tabular}[c]{@{}c@{}}Growth\\ (Time)\end{tabular}} & \textbf{\begin{tabular}[c]{@{}c@{}}Growth\\ (Mem)\end{tabular}} \\ \midrule
Rule 1 (Simple) & 463.84 & 13,132 & 2,477 & - & - \\
Rule 2 & 513.29 & 13,146 & 4,432 & +10.6\% & +0.1\% \\
Rule 3 & 926.45 & 15,396 & 4,030 & +99.7\% & +17.2\% \\
\textbf{Rule 4 (Complex)} & \textbf{937.18} & \textbf{26,132} & \textbf{5,692} & \textbf{+102.0\%} & \textbf{+99.0\%} \\ \bottomrule
\end{tabular}%
}
\end{table}

\subsubsection{Computational Latency and Real-Time Feasibility}
Beyond resource constraints, ZK also faces a critical proving-time challenge. Table~\ref{tab:zk_bottleneck} highlights that generating a proof for 1000 \textit{userOps} under Rule 4 requires over 937 seconds. As illustrated in Figure~\ref{fig:latency_all_rules}, which uses a logarithmic scale, the ZK method exhibits significant delays compared to other approaches. This high latency (over 15.6 minutes) makes the ZK approach impractical for real-time applications such as high-frequency trading or interactive gaming.

In contrast, GasLiteAA demonstrates superior performance suitable for real-time needs. As shown in Figure~\ref{fig:latency_all_rules}, it consistently maintains a sub-second latency ($\approx$ 0.1s) with only marginal sensitivity to rule complexity. By combining low on-chain gas costs with negligible off-chain latency, GasLiteAA emerges as a practical solution for supporting high-frequency, complex account abstraction scenarios efficiently.

\begin{table}[htbp]
\caption{Impact of Allocation Rule Complexity on Gas Consumption (1000 \textit{userOps})}
\label{tab:gas_complexity_comparison}
\centering
\resizebox{\linewidth}{!}{%
\begin{tabular}{lccc}
\toprule
\textbf{Allocation Rule} & \textbf{\begin{tabular}[c]{@{}c@{}}GasLiteAA\\ (Total Gas)\end{tabular}} & \textbf{\begin{tabular}[c]{@{}c@{}}Infinitism\\ (Total Gas)\end{tabular}} & \textbf{\begin{tabular}[c]{@{}c@{}}Overhead\\ (Infinitism)\end{tabular}} \\ \midrule
Rule 1 (Simple) & 82.15 M & 112.45 M & +36.9\% \\
Rule 2 (Daily+Total) & 82.15 M & 113.52 M & +38.2\% \\
Rule 3 (Interval) & 82.15 M & 134.80 M & +64.1\% \\
\textbf{Rule 4 (Dynamic)} & \textbf{82.15 M} & \textbf{158.48 M} & \textbf{+92.9\%} \\ \bottomrule
\multicolumn{4}{l}{\footnotesize * GasLiteAA costs remain constant as validation logic is offloaded to TEE.}
\end{tabular}%
}
\end{table}

\subsubsection{Impact of Rule Complexity and Resource Consumption}
We further evaluated how the complexity of validation logic affects system performance. Table~\ref{tab:gas_complexity_comparison} compares the gas consumption of GasLiteAA versus Infinitism under a fixed load of 1000 \textit{userOps} across four different rules.

The results show that GasLiteAA offers a critical advantage in cost stability. Since the complex validation logic is executed off-chain within the TEE, the on-chain gas cost for GasLiteAA remains constant ($\approx$ 82.15 M) regardless of rule complexity. In contrast, the Infinitism approach executes all logic on the EVM. Consequently, as the rule evolves from a simple check (Rule 1) to a dynamic historical calculation (Rule 4), the gas cost for Infinitism surges from 112.45 M to 158.48 M, increasing the overhead from 36.9\% to 92.9\%.

\begin{figure}[htbp]
    \centering
    \includegraphics[width=\linewidth]{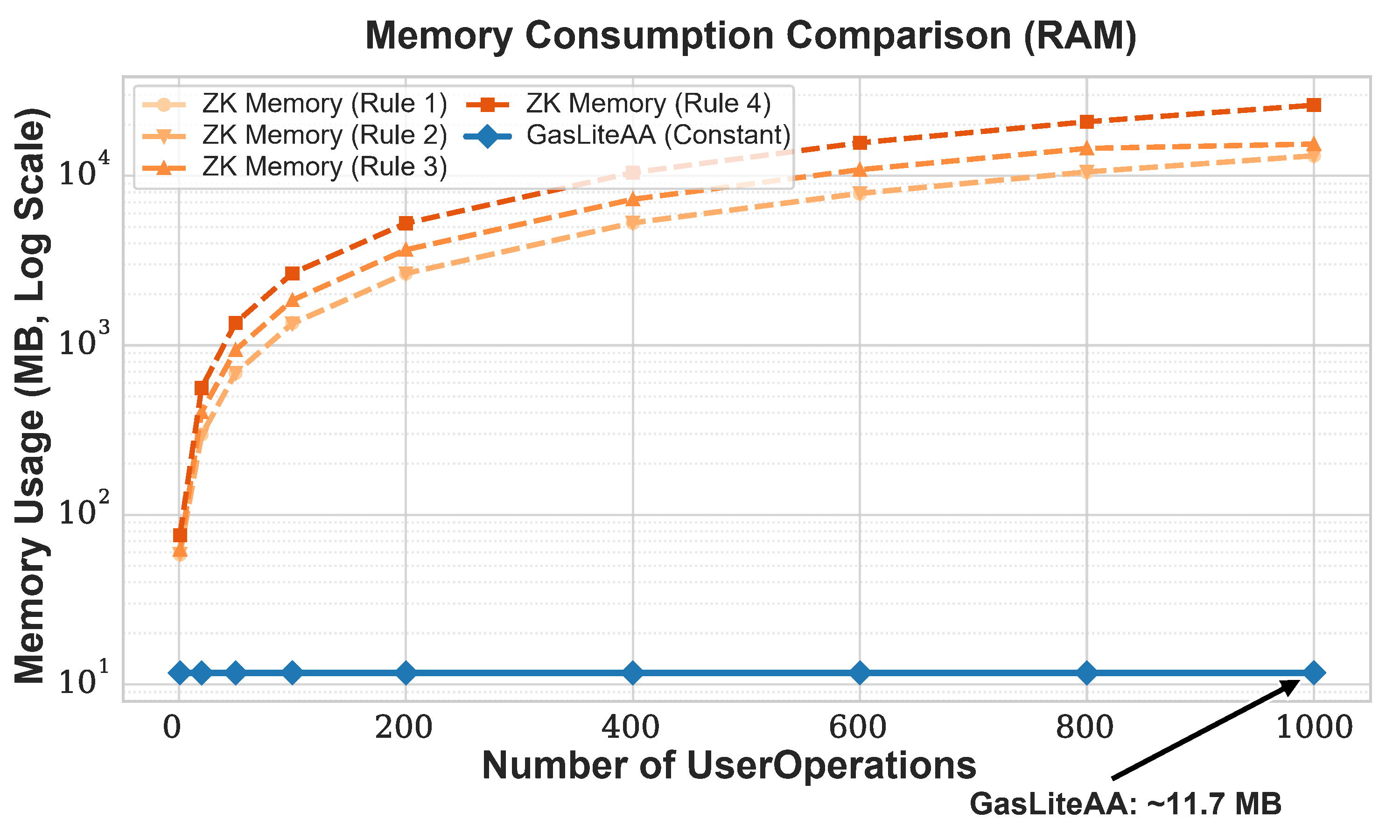}
    \caption{Memory resource consumption (RAM) comparison.}
    \label{fig:memory_comparison}
\end{figure}

\begin{figure}[htbp]
    \centering
    \includegraphics[width=\linewidth]{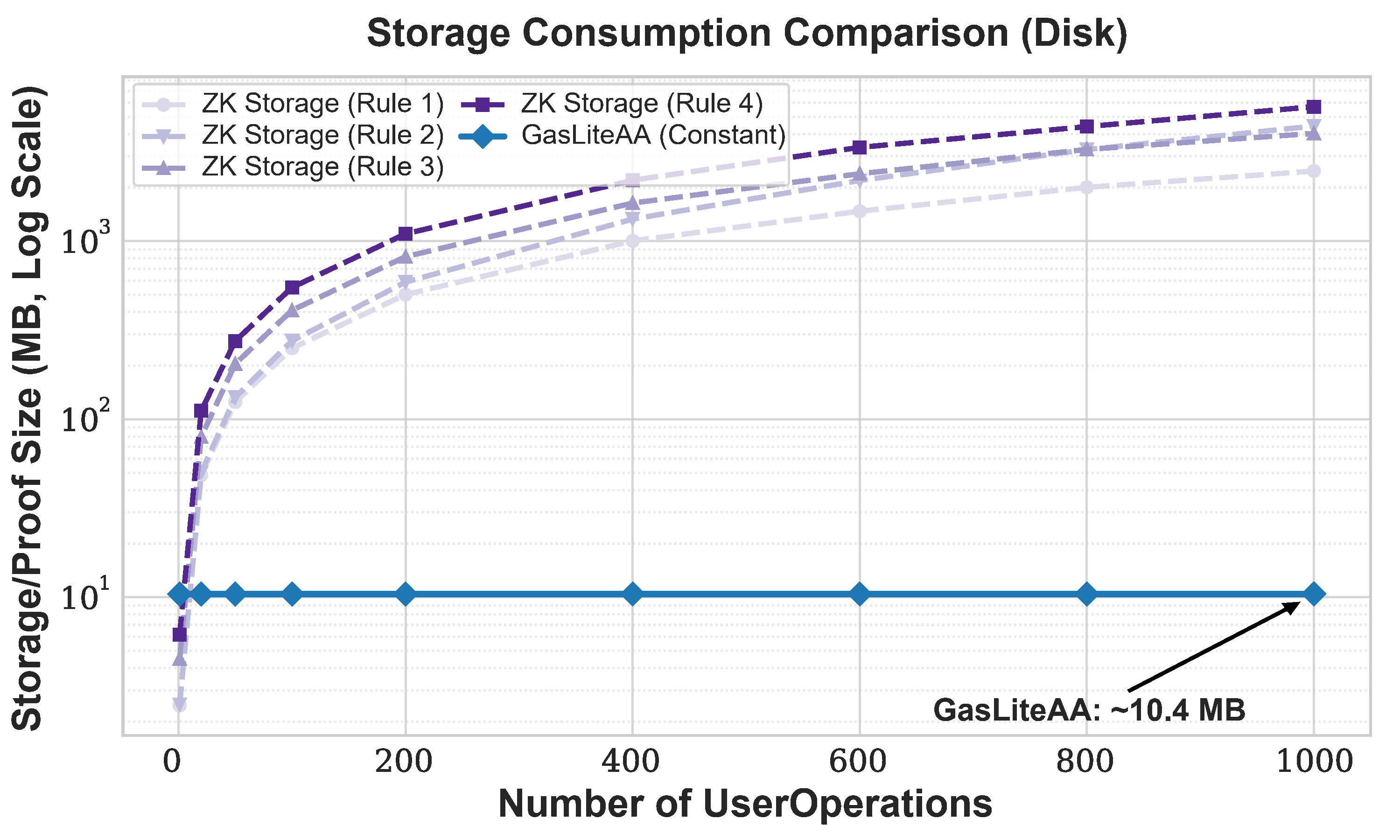}
    \caption{Storage and proof size comparison.}
    \label{fig:storage_comparison}
\end{figure}

Beyond execution time, resource overhead remains a major bottleneck for ZK proofs. Figure~\ref{fig:memory_comparison} shows that memory usage for the ZK approach surges to over 26 GB for complex rules, while GasLiteAA maintains a minimal footprint ($\approx$ 11.7 MB). Furthermore, Figure~\ref{fig:storage_comparison} highlights the storage costs associated with proof generation artifacts. The ZK approach consumes up to 5.6 GB of disk space or transmission bandwidth for complex batches, whereas GasLiteAA maintains a constant, lightweight storage requirement ($\approx$ 10.4 MB).

In sharp contrast, GasLiteAA exhibits robust stability across all metrics. The execution time within the TEE increases marginally by only 10.3\% (from 0.097s to 0.107s), and resource consumption remains negligible. This confirms that GasLiteAA is uniquely positioned to handle complex, programmable paymaster logic without sacrificing performance or demanding high-end hardware.

\section{Conclusion}
In this paper, we presented GasLiteAA, a TEE-assisted framework for ERC-4337 that offloads complex paymaster validation logic to reduce gas consumption while preserving on-chain state consistency and verifiability. We established a comprehensive system integrating components such as the \textit{userOp} Routing mechanism, Merkle tree-based state commitment, \textit{Bundler Manager}, \textit{TEE Enclave}, and \textit{AttestationOnChainVerifier}, which effectively decouples computational overhead from the Ethereum L1 mainnet. Crucially, unlike rollups that migrate execution to isolated layers, our framework enables complex stateful validation directly on L1.

We validated our approach through a robust prototype comprising approximately 28k LoC, and our evaluation demonstrates substantial efficiency gains. Specifically, in complex scenarios such as the Dynamic Daily Limit rule, traditional on-chain validation incurs a 92.9\% gas overhead compared to our optimized framework. Furthermore, ZK-based alternatives demand up to 26 GB of memory and incur a proving latency of 937.18s. Such overhead introduces perceptible user latency and increases hardware costs for \textit{Bundlers}. In contrast, GasLiteAA maintains a negligible latency of approximately 0.1s (\textbf{9,300$\times$ reduction}) and restricts memory usage to just 11.7 MB (\textbf{2,200$\times$ decrease}) compared to ZK methods. These results position GasLiteAA as a scalable, secure, and hardware-efficient solution for high-frequency account abstraction scenarios.

GasLiteAA mitigates the trade-off between programmable flexibility and gas efficiency, providing a scalable approach to AA. Future work will focus on optimizing bundling strategies, mempool routing, and more specific economic incentive mechanisms for bundlers and paymasters to further improve throughput, censorship resistance, and long-term sustainability.


\bibliographystyle{IEEEtran}

\bibliography{main}

\end{document}